# Sulfur-driven Haze Formation in Warm CO₂-rich Exoplanet Atmospheres

Chao He[1]*, Sarah M. Hörst[1,2], Nikole K. Lewis[3], Xinting Yu[1,4], Julianne I. Moses[5], Patricia McGuiggan[6], Mark S. Marley[7], Eliza M.-R. Kempton[8], Sarah E. Moran[1], Caroline V. Morley[9], & Véronique Vuitton[10]

[1] Department of Earth and Planetary Sciences, Johns Hopkins University, Baltimore, MD, USA che13@jhu.edu

[2] Space Telescope Science Institute, Baltimore, MD, USA

[3] Department of Astronomy and Carl Sagan Institute, Cornell University, Ithaca, New York 14853, USA

[4] Department of Earth and Planetary Sciences, University of California Santa Cruz, CA 95064, USA

[5] Space Science Institute, Boulder, CO, USA

[6] Department of Materials Science and Engineering, Johns Hopkins University, Baltimore, MD, USA

[7] NASA Ames Research Center, Mountain View, CA, USA

[8] Department of Astronomy, University of Maryland, College Park, MD, USA

[9] Department of Astronomy, the University of Texas at Austin, Austin, TX, USA

[10] Institut de Planétologie et d'Astrophysique de Grenoble, Université Grenoble Alpes, CNRS, Grenoble, FR

**Total pages: 32**

**The number of references: 51**

**6 figures and 1 table**

**Sulfur gases significantly affect the photochemistry of planetary atmospheres in our Solar System, and are expected to be important components in exoplanet atmospheres. However, sulfur photochemistry in the context of exoplanets is poorly understood due to a lack of chemical-kinetics information for sulfur species under relevant conditions. Here, we study the photochemical role of hydrogen sulfide ($H_2S$) in warm $CO_2$-rich exoplanet atmospheres (800 K) by carrying out laboratory simulations. We find that $H_2S$ plays a significant role in photochemistry, even when present in the atmosphere at relatively low concentrations (1.6%). It participates in both gas and solid phase chemistry, leading to the formation of other sulfur gas products ($CH_3SH/SO$, $C_2H_4S/OCS$, $SO_2/S_2$, and $CS_2$) and to an increase in solid haze particle production and compositional complexity. Our study shows that we may expect thicker haze with small particle sizes (20 to 140 nm) for warm $CO_2$-rich exoplanet atmospheres that possess $H_2S$.**

Observations[1-5] and laboratory simulations[6,7] have shown that clouds and/or hazes are likely ubiquitous in the atmospheres of exoplanets. These clouds and hazes play an important role in exoplanet atmospheres and affect the spectra of planets, therefore impacting our ability to observe their atmosphere and assess their habitability. Although a variety of atmospheric gases can condense at specific temperature and pressure conditions to form clouds, haze particles may be produced photochemically over a range of temperatures, pressures, and atmospheric compositions.[6–9] Among the various atmospheric components, sulfur gases significantly influence the photochemistry and haze formation in the atmospheres of Solar System bodies, such as Earth, Venus[10,11], Jupiter[12,13], and its moon, Io.[14] Sulfur photochemistry should also be important in exoplanet atmospheres since sulfur gases are expected to be present.[15] Photochemical modeling has considered chemical reactions of H, C, O, N, and S bearing species in the atmospheres of terrestrial and giant exoplanets, and shown that sulfur photochemistry can potentially generate sulfur hazes.[16–18] The resulting sulfur hazes could alter the transmission, thermal emission, and reflected light spectra of an exoplanet and impact observations.[19] These atmospheric models however only considered the sulfur hazes in the form of elemental sulfur or sulfuric acid, but the complex sulfur chemistry may result in other forms of sulfur-containing hazes due to the unique bonding behavior of sulfur (it has oxidation numbers of -2, 0, +2, +4 and +6). Therefore, the role of sulfur gases in photochemistry and haze formation remains unclear for exoplanet atmospheres from models alone. In addition, sulfur photochemistry is intriguing because sulfur is one of the essential elements for life on Earth and is found in several amino acids (cysteine and methionine) and some common enzymes. Understanding sulfur photochemistry may allow us to evaluate whether the photochemical sink of sulfur gases could provide a prebiotic sulfur source for life to originate extraterrestrially and examine whether or not sulfur compounds in an atmosphere can be considered as potential biosignatures on exoplanets.[20]

Laboratory experimental simulations of atmospheric photochemistry using an energy source (ultraviolet photons or cold plasma) have helped us gain valuable information about worlds in our solar system. Recent laboratory simulations have improved our understanding of photochemical processes for haze formation in atmospheres of super-Earths and mini-Neptunes.[6,7,21,22] Sulfur gases have not previously been included in these simulations except for a few that include $SO_2$ to study sulfur aerosol formation in Earth's atmosphere.[23–25] Despite the necessity of sulfur laboratory experiments for understanding observations, performing such experiments is practically challenging due to sulfur's high reactivity.

One difficulty in simulating exoplanet atmospheres is how to choose the initial gas mixture, since we do not have well-constrained atmospheric compositions of exoplanets yet. For the experiments presented here, we used a thermochemical equilibrium model of H, He, C, O, N, and S as a guide for our initial gas mixtures (see Methods)[7,15], as thermochemical equilibrium can provide a good first-order prediction of the dominant available constituents.[7] An atmosphere in thermochemical equilibrium can contain a range of molecules, such as $H_2$, $H_2O$, $N_2$, CO, $CO_2$, and $H_2S$, for which the relative abundances of the different molecules will depend on the atmospheric temperature, pressure, and bulk elemental ratios. Sulfur gases should be present in most planetary atmospheres but will tend to be more abundant in atmospheres with a higher metallicity (i.e., atmospheres with a larger abundance of heavy elements in comparison to hydrogen and helium), or, in general, in atmospheres with a higher relative abundance of sulfur in comparison with other elements.[15] Fig. 1 shows our initial gas mixture, based on equilibrium compositions for 10,000× solar metallicity at 800 K and 1 mbar. Our gas mixture has a high mean molecular weight of ~32 μ, which is relevant to the atmospheres of sub-Neptunes, Super-Earths, and Earths. We run the experiments at 800 K and 4.5 mbar, which is consistent with the values used in the model calculations. There are currently ~20 Transiting Exoplanet Survey Satellite (TESS)[26] objects of interest less than 2.0 Earth radii with equilibrium temperatures between 700 and 900 K. Additionally, there are a host of known observed exoplanets which occupy this temperature regime, such as GJ 436 b[27] and HD 97658 b[28], which have muted water features in transmission observations.[2,3] The muted features could be potentially caused by photochemical hazes. However, the photochemistry at this temperature range has not been experimentally studied. The calculated gas composition serves as one example of diverse high-metallicity atmospheres rather than representing one specific exoplanet, but provides insight into the sulfur photochemistry in $CO_2$-rich exoplanet atmospheres.

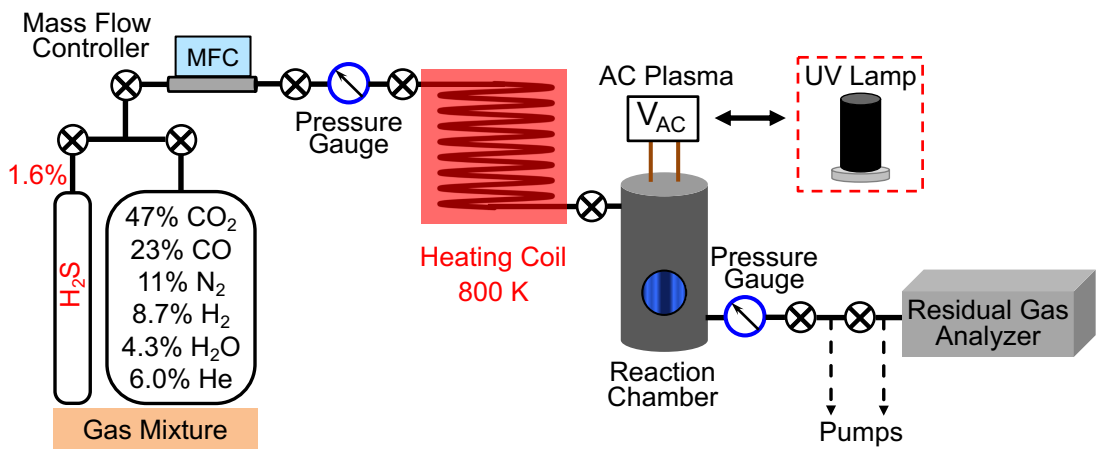

**Fig. 1 Simplified schematic of the PHAZER chamber for current work.** In this study, we use one of two energy sources to initiate the photochemistry in the chamber, either a cold plasma generated by an AC glow discharge or FUV photons produced by a hydrogen lamp. The 800 K, 10,000× solar metallicity gas mixture used here was calculated in the same manner as He et al. (2018)[6] and Hörst et al. (2018).[7] The current investigation is an extension of those studies but more specifically focuses on the sulfur photochemistry in exoplanet atmospheres.

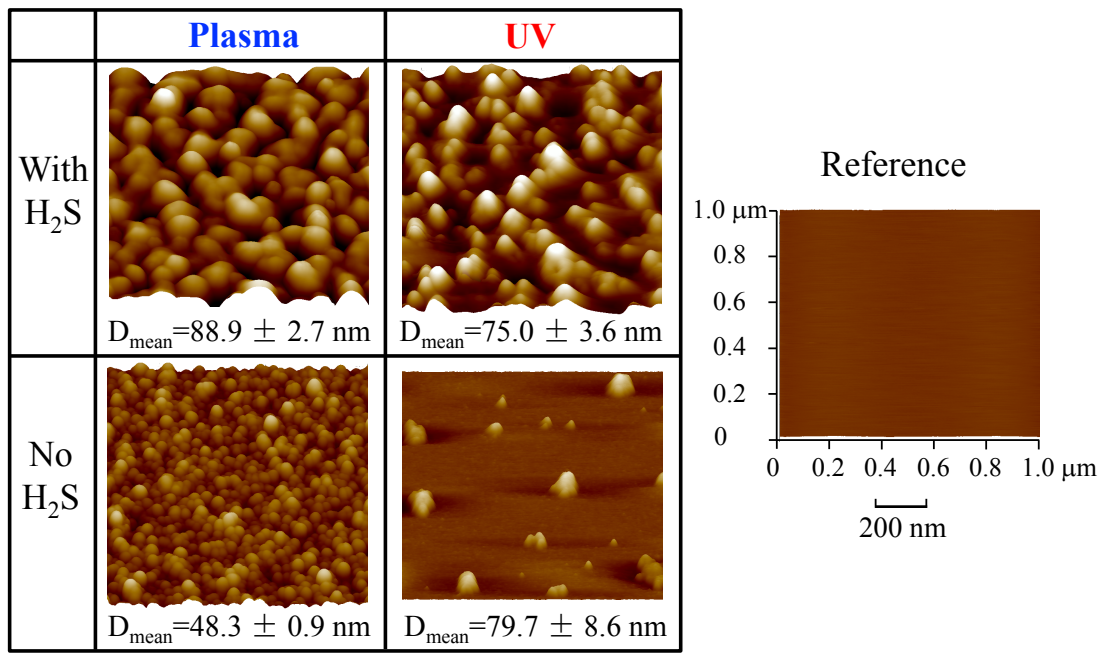

**Fig. 2 AFM images of the particles on mica substrates.** The scanning area is 1 μm × 1 μm for each sample. "Reference" is the AFM image of the disk from the reference experiment (the gas mixture with $H_2S$ at 800 K without plasma or UV exposure). The mean diameters ($D_{mean}$, nm) and the associated uncertainties of the haze particles are shown under the images.

## Results

We performed two sets of experiments, one including $H_2S$, and the other without, using the Planetary Haze Research (PHAZER) chamber (Fig. 1).[29] We used two energy sources because they simulate different processes in planetary atmospheres (AC plasma mimic energetic planetary upper atmospheres and UV photons simulate stellar UV radiation) and may serve as a proxy for varying energy fluxes in diverse stellar environments.[6,7,22] In the chamber, the heated gas mixtures (800 K) were exposed to one of two energy sources (AC plasma or FUV photons). The gas mixtures flowing out of the chamber were analyzed with a mass spectrometer (see Methods). After a 72 h continuous run, the solid samples (powders from the wall and/or films on mica substrates) were collected in a dry-nitrogen glove box. For both gas mixtures, the plasma experiments produced enough solid particles to collect and weigh, while the UV experiments only resulted in thin films. With an atomic force microscope (AFM), we characterized the films on the mica substrates from our experiments (See Methods). As shown in Fig. 2, the AFM image from the "Reference" experiment (the gas mixture with $H_2S$ at 800 K without plasma or UV exposure) shows that the mica surface is smooth and clean, indicating that merely heating the gas mixture does not generate particles. In contrast, the AFM images reveal spherical particles on the substrates from the plasma and UV experiments, confirming that both the plasma and UV experiments produce solid particles for the gas mixtures whether or not $H_2S$ was included (Fig. 2). However, the number and the size of the particles are different in the four cases (plasma-$H_2S$, plasma-no-$H_2S$, UV-$H_2S$, and UV-no-$H_2S$). In general, the plasma-$H_2S$ experiment produces bigger particles ($D_{mean}$=88.9 nm) than other cases, while the plasma-no-$H_2S$ experiment produces the most but relatively small particles ($D_{mean}$=48.3 nm). Both UV experiments produce particles in a wider size range, but the UV-no-$H_2S$ produces the fewest particles among the four cases.

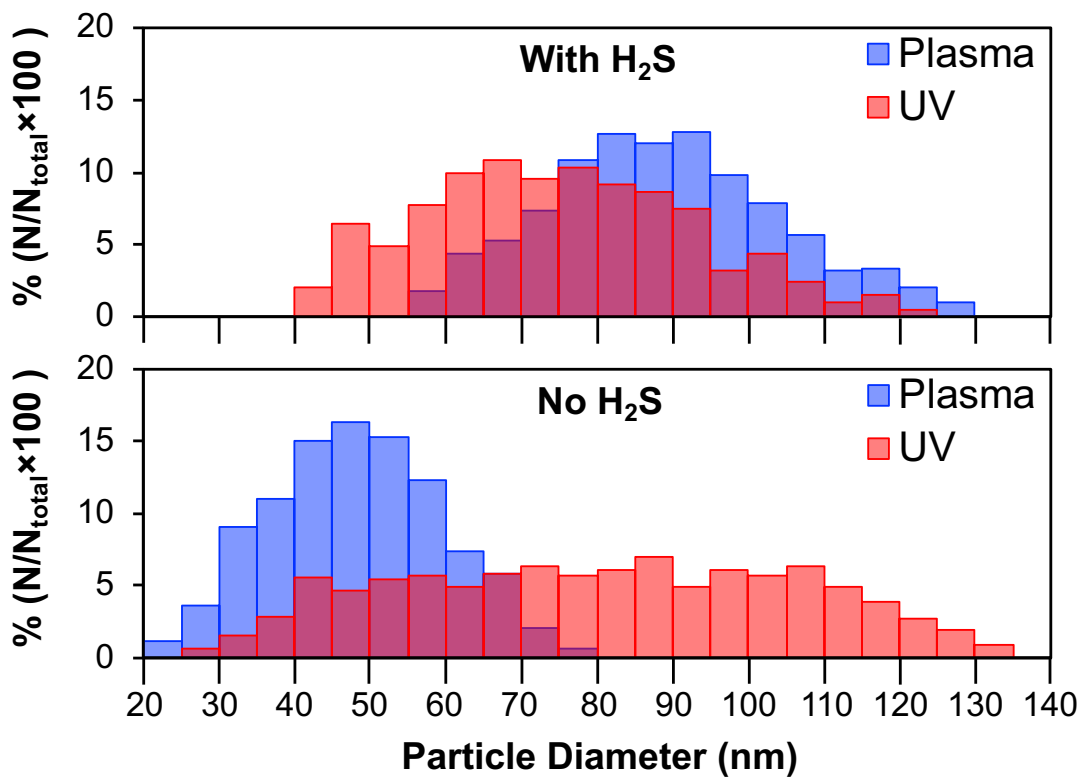

**Fig. 3 Size distribution of the haze particles formed in plasma (blue) and UV (red) experiments, with H₂S (top), or no H₂S (bottom).** The particles generated in the UV experiments have a broader size range, while the particle sizes are more uniform for the plasma experiments. For the plasma, particles are uniformly large with H₂S and uniformly small without H₂S.

The particle size distribution for the four experiments measured from the AFM is shown in Figure 3 with 5 nm size bins (see Methods). It shows that the particles from the plasma-H₂S experiment are in the range of 55 to 130 nm, while those from the plasma-no-H₂S experiment are smaller (20 to 80 nm). The particles from the UV experiments have wider size distributions, 40 to 125 nm for the UV-H₂S experiment, and 25 to 135 nm for the UV-no-H₂S case. Figure 3 shows that the UV experiments generate particles in a broader size range, while the particle sizes are more uniform for the plasma experiments (either uniformly large or uniformly small). The UV experiment has a lower production rate than the plasma experiment due to lower energy density[22], which likely causes the UV experiments to have fewer nucleation centers and a slower particle growth rate. Thus, gas-solid heterogenous reactions are more likely to happen in the UV experiments and lead to broader size range. Besides the energy density, the energy type could also affect the size range of the particles. In our experiments, the haze particles are in the range of 20 to 140 nm. This size range is similar to previous experiments[8,21], and lies in the Rayleigh scattering regime for visible and infrared (IR) photons, in which small particles (compared to the photon wavelength) will scatter short wavelengths more efficiently. If such small particles are present in exoplanet atmospheres, they would drastically impact the spectra of an exoplanet[30], therefore affecting the observations of exoplanet atmospheres by current[31] and future telescopes.[9,19]

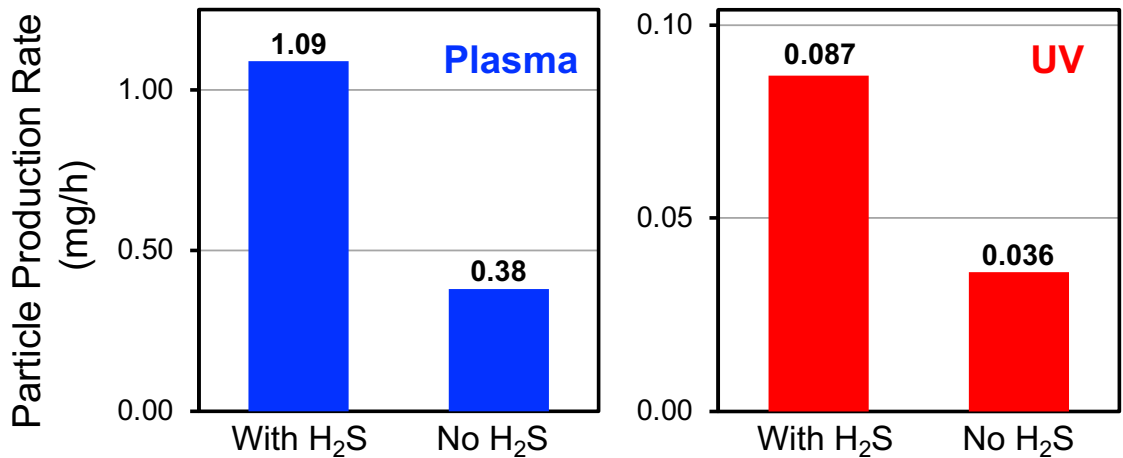

**Fig. 4 Haze particle production rate in the plasma (blue) and UV (red) experiments.** Note that the scales are different between the two panels. The plasma experiments have higher haze production rates than the UV experiments. Including $H_2S$ in the initial gas

mixture increased the haze particle production rate by ~3 times for both energy sources. The production rates in the plasma experiments are calculated by the weighing method and the uncertainty is 0.00015 mg h$^{-1}$. For the UV experiments, the uncertainties of the estimated production rates (from particle size distribution) are 2.4% (no $H_2S$) and 3.0% (with $H_2S$). We also estimated the production rates for the plasma experiments from the particle size distribution, which are 0.13 mg h$^{-1}$ with 6.2% uncertainty (no $H_2S$) and 0.25 mg h$^{-1}$ with 2.7% uncertainty (with $H_2S$). These values are lower than those calculated from the weighing method because the plasma experiments produced enough particles to form multiple layers on the surface. Note that the production rates calculated by both methods are likely to be a lower limit. For the weighing method, it is not possible to collect all the solid sample on the wall. For the production rate estimated from the particle size distribution, there could be multiple layers of particles on the substrate.

Fig. 4 shows the production rates of the solid particles in the plasma and UV experiments. For the UV cases, we calculated the production rates based on the particle size distribution derived from the AFM images (the reported production rate for the UV experiment is a lower limit, see Methods). For the plasma experiments, the production rates were calculated by weighting the collected solids (see Methods). Fig. 4 shows that, whether or not we include $H_2S$ in the initial gas mixtures, the production rates of the solid particles are approximately 10 times higher in the plasma experiments than in the UV experiments. This could be caused by higher energy density of the plasma as compared to the UV radiation.[22] Note that plasma and UV radiation are two different energy sources and they are used to simulate different processes in planetary atmospheres. The plasma could mimic electrical activities and/or charged particles in planetary upper atmospheres while the UV photons (110 to 400 nm) produced by the lamp are used to simulate stellar UV radiation.[32] In our experiments, the photons are unable to directly dissociate very stable molecules such as $N_2$ or CO, but the photons (< 317 nm) can break down $H_2S$ directly. The plasma we used is very energetic and can destroy very stable molecules including $N_2$ or CO.

For the same energy source, the experiments which include $H_2S$ in the initial gas mixture produce about 3 times more solid particles than those without $H_2S$. This result indicates that the addition of $H_2S$ in the gas mixture, even at a low level (1.6%), substantially promotes the formation of haze particles regardless of the type of energy input (plasma or UV radiation). Due to the addition of $H_2S$ and the higher energy density of plasma, the plasma-$H_2S$ experiment produces larger particles with a higher production rate, resulting in the highest haze mass of all experiments considered here.

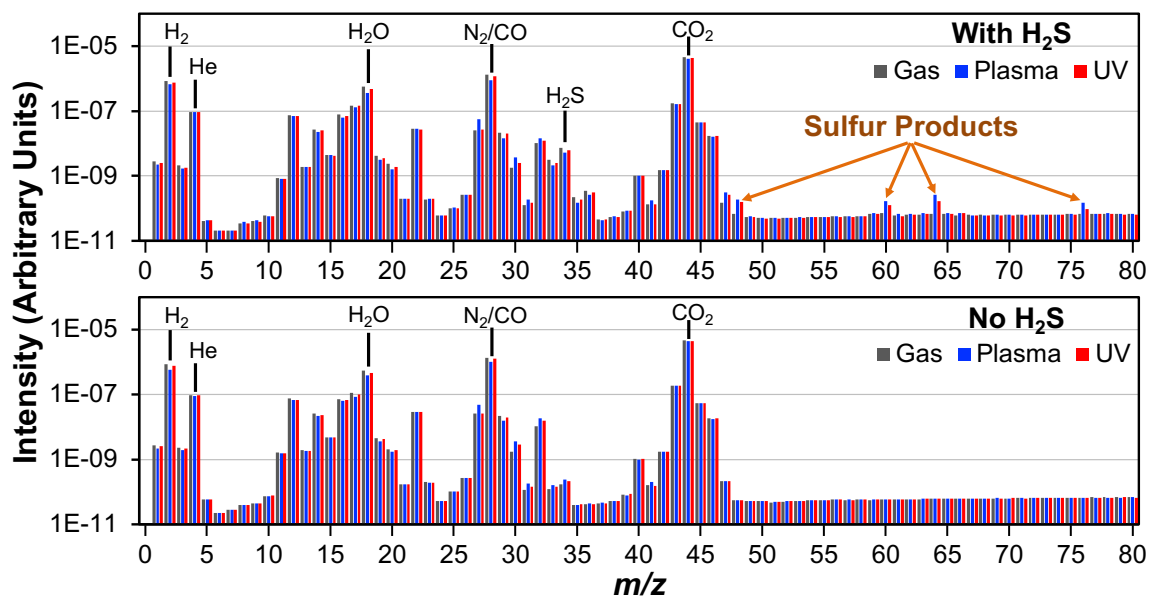

**Fig. 5 The mass spectra of the gas mixtures with $H_2S$ (top) and without $H_2S$ (bottom): initial gas mixture (gray), plasma on (blue), and UV on (red).** The initial gas compositions are labeled near the peaks and have an intensity decrease under energy

exposure. The increased peaks are associated with new gas phase products of photochemistry. The mass spectra have a background level at ~6×10$^{-11}$ intensity.

**Table 1. Assignments and the relative intensity change of the changed molecular peaks in the mass spectra (Fig. 5).**

| Peak (m/z) | Species | With $H_2S$ | | No $H_2S$ | |
|---|---|---|---|---|---|
| | | Plasma | UV | Plasma | UV |
| 2 | $H_2$ | ↓ 1.8E-7 | ↓ 6.7E-8 | ↓ 2.7E-7 | ↓ 7.8E-8 |
| 18 | $H_2O$ | ↓ 1.9E-7 | ↓ 8.7E-8 | ↓ 1.4E-7 | ↓ 8.5E-8 |
| 28 | $N_2/CO$ | ↓ 4.2E-7 | ↓ 1.6E-7 | ↓ 3.1E-7 | ↓ 7.1E-8 |
| 34 | $H_2S$ | ↓ 2.2E-9 | ↓ 1.2E-9 | — | — |
| 44 | $CO_2$ | ↓ 3.2E-7 | ↓ 1.1E-7 | ↓ 1.7E-7 | ↓ 3.9E-8 |
| 27 | HCN | ↑ 3.2E-8 | ↑ 1.3E-9 | ↑ 2.4E-8 | ↑ 1.2E-9 |
| 30 | $C_2H_6/HCHO/NO$ | ↑ 1.9E-9 | ↑ 7.2E-10 | ↑ 2.0E-9 | ↑ 1.3E-9 |
| 31 | $CH_3NH_2$ | ↑ 5.8E-11 | ↑ 2.6E-11 | ↑ 7.2E-11 | ↑ 3.5E-11 |
| 32 | $O_2/CH_3OH$ | ↑ 3.9E-9 | ↑ 1.4E-9 | ↑ 7.8E-9 | ↑ 5.2E-9 |
| 41 | $CH_3CN$ | ↑ 4.6E-11 | — | ↑ 5.0E-11 | — |
| 48 | $CH_3SH/SO$ | ↑ 1.2E-10 | ↑ 9.1E-11 | — | — |
| 60 | $OCS/C_2H_4S$ | ↑ 9.5E-11 | ↑ 5.6E-11 | — | — |
| 64 | $SO_2/S_2$ | ↑ 1.9E-10 | ↑ 9.6E-11 | — | — |
| 76 | $CS_2$ | ↑ 8.1E-11 | ↑ 2.5E-11 | — | — |

↓: Decrease; ↑: Increase; —: No change.

Figure 5 shows the mass spectra of the gas mixtures before and after turning on the energy source. The molecules in the initial gas mixtures are labeled on the spectra. The top panel includes $H_2S$ in the initial gas mixtures while the bottom panel does not (Fig. 5). By comparing the mass spectra, we notice that the intensity of some peaks changes after turning on the energy source. The decreased peaks are due to the destruction of the initial gas molecules while the ones that increase are new gas products, as a result of the chemistry initiated by the energy source. Table 1 lists the most probable species corresponding to the peaks that changed. Some peaks have several possibilities because there are species with identical nominal mass that could not be resolved in our unit-resolution mass spectra. Except helium (inert gas), all other molecules in the initial gas mixtures ($H_2$, $H_2O$, $N_2/CO$, $CO_2$, and $H_2S$) exhibit a decrease in intensity under energy exposure (Fig. 5).

The increased peaks are associated with the newly-formed gas products, but the species and the quantity vary between different experiments. For the experiments without $H_2S$, the molecular peaks that increased are 27, 30, 31, and 32 amu in the UV case. In the plasma-no-$H_2S$ experiment, the same peaks increase plus one at 41 amu. These peaks are hydrogen cyanide (HCN, 27 amu), ethane/formaldehyde/nitric oxide ($C_2H_6$/HCHO/NO, 30 amu), methylamine ($CH_3NH_2$, 31 amu), oxygen/methanol ($O_2$/$CH_3OH$, 32 amu, but $O_2$ should be more abundant under these conditions), and acetonitrile ($CH_3CN$, 41 amu), respectively. The increase of mass peak 30 is probably contributed by HCHO/NO rather than $C_2H_6$ because there is no $CH_4$ initially in the system and the initial gas mixtures are relatively oxidized, but we cannot completely rule out $C_2H_6$. Here, $O_2$ is produced abiotically by photochemistry, not as oxygen in Earth's atmosphere which is produced by photosynthesis by plants/bacteria. Therefore, our result reiterates our previous experimental results[22] that we should rule out the abiotic sources before considering $O_2$ as potential biosignature for other planets.[33,34]

For the experiments including $H_2S$ (at 1.6%), there are 4 additional peaks that increased (48, 60, 64, and 76 amu) in both the plasma and UV cases. These peaks most likely correspond to sulfur species since they are produced after introducing $H_2S$ into the initial gas mixtures, and all four peaks match with sulfur-containing molecules. These include methanethiol/sulfur monoxide ($CH_3SH/SO$, 48 amu), carbonyl sulfide/thiirane ($OCS/C_2H_4S$, 60 amu), sulfur dioxide/disulfur ($SO_2/S_2$, 64 amu), and carbon disulfide ($CS_2$, 76 amu). Considering the complexity of sulfur photochemistry[18,20], it is possible that both $CH_3SH$ and SO are responsible for the increase of mass 48, both OCS and $C_2H_4S$ for mass 60, and both $SO_2$ and $S_2$ for mass 64. Although the unit-mass resolution does not allow differentiating the candidate molecules, the fragmentation patterns could provide some additional clues to identify some of the products. For example, the mass spectra show an increase of the peak at 47 amu in the experiments including $H_2S$ with both energy inputs, and the intensity is higher than the one at 48 amu. This is consistent with the fragmentation pattern of $CH_3SH$. It is likely that the peak at m/z 48 could be attributed to $CH_3SH$ rather than SO. Regarding the two candidates ($C_2H_4S$ and OCS) for the peak at 60 amu, OCS has a major peak at m/z 60 and another important fragment at 32 while $C_2H_4S$ has a major fragment at m/z 45 and important fragments at m/z 59 and 60. The mass spectra in Figure 5 appear more compatible with OCS than $C_2H_4S$. However, the unit-resolution mass spectra do not allow us to determine the ratios of two products for each peak.

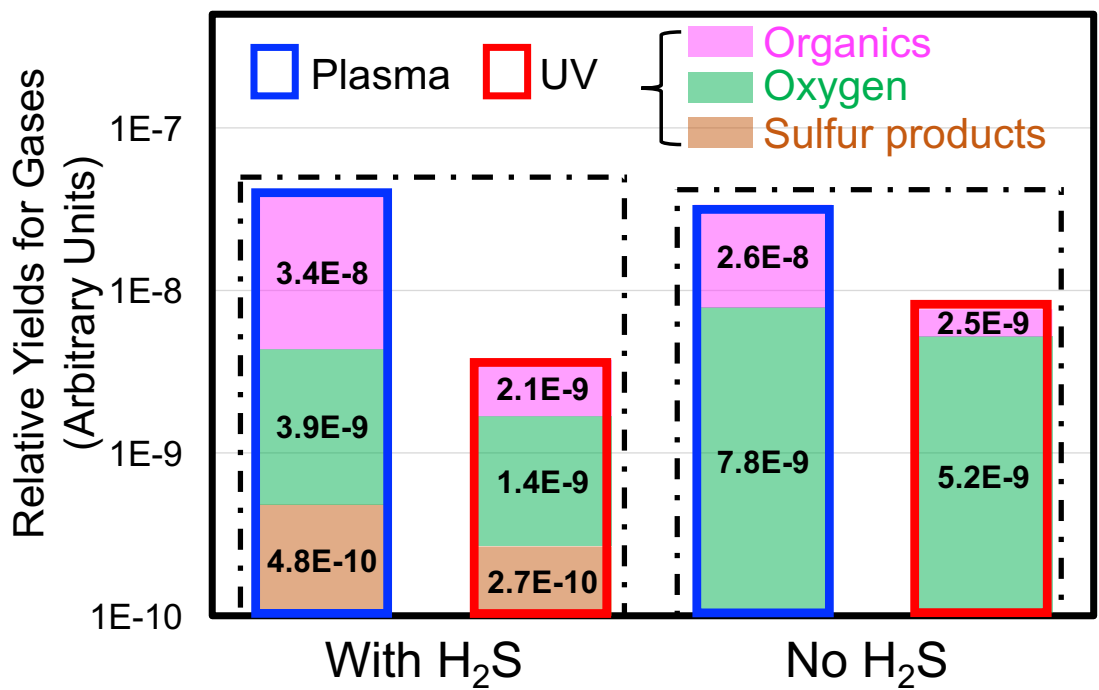

**Fig. 6 Relative yields of the gas products in the plasma (blue) and UV (red) experiments.** Organics refers to organic gas products that contain C, H, O, and N; while sulfur products refer to both organic and inorganic gas products that contain sulfur. The uncertainties (due to the measurement error of the mass spectra) for the relative yields are less than 1%. The plasma experiments have higher gas product yields than the UV experiments. The experiments including $H_2S$ produce new sulfur gas products for both energy sources.

To better compare the gas products between experiments, we grouped the gas products in three categories (organics, $O_2$, and sulfur products that contain sulfur but can be either inorganic or organic) and calculated the relative yield for each category (see Methods) as shown in Fig. 6. The gas phase products are similar for the same initial gas mixture with two different energy sources (plasma or UV), but the abundance of these products differs depending on energy source. For the same gas mixture, the total yields of the new products from the plasma experiments are higher than from the UV experiments (~10 times higher for the experiments with $H_2S$, ~5 times higher for the no-$H_2S$ experiments). The higher-yield trend (plasma versus UV) is true for each category of the products, which reflects the higher energy density of the plasma.[22] The estimated energy densities of the two different energy sources (170 $W/m^2$ for the AC glow discharge, and 36 $W/m^2$ for the UV lamp) are higher than a hypothetical small (2 Earth radius), warm (800 K) exoplanet (10 $W/m^2$) around a given host M-star (3000 K). Therefore, a 72-hour of exposure to the AC glow discharge in our experiment roughly corresponds to 51 days (72-hour of exposure to the UV lamp corresponds to 10 days) of UV irradiation from the host M-star. Comparing between the experiments with two different initial gas mixtures (including $H_2S$ or not), we find that the total yields of the new products are comparable for the experiments with same energy source. In particular, the yields of the organic products are similar. The sulfur products are only generated in the experiments with gas mixtures including $H_2S$. The yield of oxygen in the experiment without $H_2S$ is higher than that in the experiment with $H_2S$. The initial gas mixture is $CO_2$-rich and relatively oxidized, while $H_2S$ is reducing. Introducing $H_2S$ into the gas mixture changes the total oxidizing environment of the system, thus reducing the production of oxygen.

The initial gas mixture contains $CO_2$, CO, $N_2$, $H_2$, and $H_2O$, with the addition of $H_2S$ for one set of experiments. The energy input (plasma or UV radiation) initiates chemical reactions in the gas mixtures, and further reactions lead to the creation of new gas products and eventually the solid particles. Gas composition changes, along with the haze production rate, can shed light on the chemical pathways responsible for the haze formation. For instance, the organic gas product HCN is observed in all of our experiments, although there are no organic molecules (e.g., $CH_4$) in the initial gas mixtures. HCN can react with itself and many other species, and further reactions can form larger molecules and organic haze particles. The carbon in the observed organic molecules must come from $CO_2$ or CO in the initial mixtures. CO serves as a better carbon source for organics than $CO_2$ as it is more efficient at producing atomic carbon from dissociation of CO as demonstrated in previous studies.[9,22] Modeling studies[15,35] have shown that HCN can be photochemically produced in $CO/N_2/H_2O$ or $CO_2/N_2/H_2O$ atmospheres. Mass peak 30 could be NO, HCHO, and/or $C_2H_6$. NO was suggested to form under similar conditions by a previous modeling study[15], and photochemical production of HCHO in $CO/H_2O$ or $CO_2/H_2O$ has also been reported experimentally.[36,37] NO could contribute to nitrogen incorporation into the haze particles, while $C_2H_6$ and/or HCHO could contribute to organic haze formation. HCN and HCHO can further react to produce many other organic species such as amines, alcohols, amides, aldehydes, ketones, and more complex organics. Therefore, they can serve as key precursors for producing more complex compounds and haze particles, and should be treated as important atmospheric indicators for complex photochemistry when observing exoplanet atmospheres. Additionally, they are also very crucial prebiotic precursors for producing sugars, amino acids, and nucleobases.[38,39]

After introducing $H_2S$ into the gas mixture, several sulfur products are detected besides the organic molecules and $O_2$, including $CH_3SH/SO$, $C_2H_4S/OCS$, $SO_2/S_2$, and $CS_2$. The variety of the sulfur products suggests that $H_2S$ interacts with other molecules in the system and induces rich chemical reactions. The production and loss mechanisms for sulfur species under the conditions of our experiments are uncertain, due to a lack of laboratory or theoretical rate-coefficient data for many reactions of interest. However, we can look to photochemical modeling studies of the Earth's atmosphere[40], early Earth[41], Venus[10,42], the Comet Shoemaker-Levy 9 impacts with Jupiter[12], Io[14], Titan[43], and exoplanets[16–18,20] to estimate the important pathways.

The $H_2S$ in our reaction chamber can be dissociated directly by plasma or UV photons with wavelengths less than 317 nm (R1), or the $H_2S$ can be destroyed through reactions (R2, R3) with OH and H produced from water photolysis. These processes produce SH radicals that can react with various constituents in the gas phase. The SH radicals can further react with other SH radicals to form atomic S (R4), react with atomic oxygen from $CO_2$ photolysis (in the plasma and UV experiments) or CO photolysis (in the plasma experiments) to form SO (R5), react with atomic carbon from CO photolysis (in the plasma experiments) to form CS (R6), or react with atomic S to form $S_2$ (R7).[16–18,20,40] These products can then further react, leading to a rich and complex sulfur chemistry. For example, $S_2$ can polymerize to $S_8$ (R8 and R9) as suggested by Zahnle et al.[17,18]

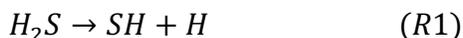
$$H_2S \rightarrow SH + H \qquad (R1)$$

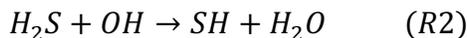
$$H_2S + OH \rightarrow SH + H_2O \qquad (R2)$$

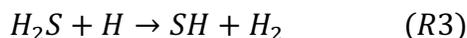
$$H_2S + H \rightarrow SH + H_2 \qquad (R3)$$

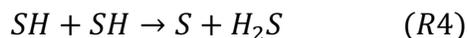
$$SH + SH \rightarrow S + H_2S \qquad (R4)$$

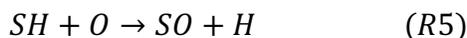
$$SH + O \rightarrow SO + H \qquad (R5)$$

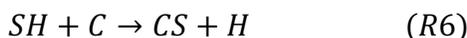
$$SH + C \rightarrow CS + H \qquad (R6)$$

$$SH + S \rightarrow S_2 + H \qquad (R7)$$

$$S_2 + S_2 \rightarrow S_4 \qquad (R8)$$

$$S_4 + S_4 \rightarrow S_8 \qquad (R9)$$

At the temperature, pressure, and bulk elemental abundances in our reaction chamber, OCS (60 amu) is one of the most important sulfur-bearing species in thermochemical equilibrium. Multiple kinetic pathways likely contribute to the formation of OCS, such as reactions between OH radicals and CS (R10), reactions between S and HCO (R11), and the termolecular addition of CO and S (R12).[18,20,40] OCS plays a significant role in the sulfur reaction network, which can participate in a range of gas phase reactions and generate various sulfur products.[40,44] In addition, OCS is considered as an important prebiotic molecule because it can catalyze the formation of peptides from amino acids.[45]

$$CS + OH \rightarrow OCS + H \qquad (R10)$$

$$S + HCO \rightarrow OCS + H \qquad (R11)$$

$$S + CO + M \rightarrow OCS + M \qquad (R12)$$

These newly formed sulfur-containing molecules in the gas phase indicate sulfur is likely also incorporated into the solid haze particles, since $H_2S$ and its resulting sulfur gas products are highly reactive photochemically. The addition of $H_2S$ not only changes the total oxidizing/reducing environment of the gas mixture, but also induces rich photochemistry and leads to the formation of new gas products and an increase in the haze production rate.

**Discussion**

Our experiments that include $H_2S$ produce several sulfur-containing gases. Some of these gases, such as $CH_3SH$, OCS, and $CS_2$, have been suggested as potential biosignatures exoplanetary atmospheres[20,46], because they are usually biogenic products on Earth. However, our result shows that they can be produced abiotically through sulfur photochemistry in exoplanet atmospheres. The possibility that these sulfur gases could act as a "false positive" for biosignatures should be reviewed carefully.

The inclusion of $H_2S$ increases the haze production rate by a factor of ~3 for both energy sources (plasma or UV). The composition of the solid haze particles can be inferred from the gas products. In the no-$H_2S$ experiments, the organic precursor, HCN, in the gas phase suggest the solid particles should be mainly organic compounds. The particles from the sulfur experiments could include organic compounds, sulfur polymers, and organosulfur compounds. This complexity is due to the presence of new sulfur gas products (such as OCS, $S_2$, $CS_2$, $CH_3SH$, and/or $C_2H_4S$) with the organic precursors in the gas phase. Models predict that sulfur chemistry produces elemental sulfur ($S_8$) hazes in $H_2$-dominated exoplanet atmospheres[16,18], or elemental sulfur ($S_8$) and/or sulfuric acid ($H_2SO_4$) hazes in oxidized exoplanet atmospheres ($N_2$- or $CO_2$-dominated)[16], but condensation of elemental sulfur ($S_8$) is not expected at the 800 K of our experiments. However, sulfur tends to form longer chains or polymers, which might become refractory enough to form solids in our experiments despite the high temperature (800 K). Sulfur species may also be incorporated into other organic molecules produced photochemically. If we assume all the decomposed $H_2S$ ends up in the solid particles as sulfur polymers, it will only account for 3.6% of the solid particles by mass in the plasma experiment, or 18.2% in the UV experiment (This ratio should be even lower because part of the decomposed $H_2S$ forms sulfur products in the gas phase). Therefore, the majority of the solid particles formed in the sulfur experiments are organic compounds, either non-sulfur-containing or organosulfur compounds.

The photochemical formation of organosulfur compounds may provide a potential chemical pathway for prebiotic evolution.[47] Organosulfur compounds, such as methionine and cysteine (two of the proteinogenic amino acids that contain sulfur), are important for the origin of life on Earth. A previous study showed that an electric discharge on a model prebiotic atmosphere ($CH_4$, $NH_3$, $CO_2$, and $H_2S$) produced methionine and other organosulfur compounds.[47] The organosulfur compounds produced in our experiments could also contain species that are of biochemical significance. Analysis of the solid samples is underway to determine their compositions. Although our experiments were conducted at 800 K, similar sulfur chemistry could happen at lower temperatures relevant to prebiotic chemistry because sulfur species are very reactive.

The enhancement of organic haze production in our experiments by $H_2S$ and its photochemical gas products is consistent with a recent modeling study.[48] In this study, Arney et al.[48] found that sulfur photochemistry acts as a major source of free radicals that can drive organic haze formation, although they focused on the photochemistry at an Earth-like temperature (~300 K). Most exoplanet haze studies (for example, Morley et al. 2015[9]) only consider simple organic chemistry in calculations when they estimate the haze production rates using photochemical modeling. Our experimental simulations show that other species, such as sulfur, participate in photochemistry and strongly affect haze production in exoplanet atmospheres. Sulfur species have distinctive optical properties; for instance, all known solid sulfur allotropes and most of organosulfur compounds have UV/blue light absorptions.[49,50] Therefore, the sulfur-incorporated organic haze particles may substantially shape the spectra of exoplanets observed with current and future facilities, such as the Hubble Space Telescope, the James Webb Space Telescope, and beyond. Further work is required to understand haze formation pathways and quantify their impacts on current and future observations.

**Data availability**

The data that support the plots within this paper and other findings of this study are available from the corresponding author upon reasonable request.

**Additional Information**

Correspondence and requests for materials should be addressed to C.He.

**Acknowledgements**

This work was supported by the NASA Astrophysics Research and Analysis Program NNX17AI87G. X. Yu is supported by a 51 Pegasi b Fellowship. S. E. Moran is supported by NASA Earth and Space Science Fellowship 80NSSC18K1109.

**Author contributions**

C.H., S.M.H., N.K.L., M.S.M. and J.I.M. conceived the study. J.I.M. calculated the starting gas mixtures. C.H. carried out the experiments and MS measurements. C.H. and X.Y. performed the AFM measurements. C.H. conducted the data analysis and prepared the manuscript. All authors participated in discussions regarding interpretation of the results and edited the manuscript.

**Competing interests**

The authors declare no competing interests.

**Methods**

**Experimental procedure.** We carried out the experiments using the PHAZER setup (Fig. 1) at Johns Hopkins University.[29] We used the thermochemical-equilibrium model of Moses et al. (2013)[15] to guide the choice of our initial gas mixtures. We included the elements H, He, C, N, O, and S in the equilibrium calculation for the case of 10,000× solar metallicity at 800 K and 1 mbar, selected gases with a calculated abundance of ≥1% (for a manageable level of experimental complexity), and renormalized the remaining mixing ratios (Fig. 1). This gas mixture is an extension of the matrix in our previous studies.[8,9,21,22] Similar as in previous studies, we also ran experiments for a range of metallicities (100x, 1000x, 10000x) at 800 K. The calculated gas mixture for the 10000x case is the only gas mixture that contains >1% of a sulfur species (1.6% of $H_2S$). With $H_2S$ in the gas mixture, we can investigate the photochemical role of sulfur gas in exoplanet atmospheres. The rest of results will be reported in a separate paper. The initial gas mixtures were prepared in the ratios as shown in Fig. 1 with high-purity gases ($CO_2$-99.999%, CO-99.99%, $N_2$-99.9997%, $H_2$-99.9999%, He-99.9995%, and $H_2S$-99.5%) and HPLC Grade water (Fisher Chemical). The premixed gas mixtures were flowed through a custom heating coil (15-meter stainless steel coil) and heated to 800 K. The heated gas mixtures were then flowed into a stainless-steel reaction chamber where they were exposed to one of two energy sources (cold plasma generated by an AC glow discharge or FUV photons produced by a hydrogen lamp). The gas flow rate was controlled by a mass flow controller (MKS Instruments) at a rate of 10 standard cubic centimeters per minute, which maintained a consistent pressure in the chamber at 4.5 mbar. The gas mixtures that flowed out of the chamber were measured with a Residual Gas Analyzer (RGA, Stanford Research Systems), and the produced solid particles remained inside of the chamber and deposited on the wall of the chamber and mica substrates placed in the chamber. After a 72-hour continuous run, the energy source was turned off and the chamber cooled down to ambient temperature. The chamber was pumped down to $10^{-3}$ mbar and further kept under vacuum for 48 hr to remove the remaining volatiles. The chamber was transferred to a dry (<0.1 ppm $H_2O$), oxygen free (<0.1 ppm $O_2$) $N_2$ glove box (I-lab 2GB; Inert Technology Inc.) where the solid samples (powders from the wall and films on mica substrates) were collected. The powders were weighed (Sartorius Entris 224-1S with a standard deviation of 0.1 mg) in the glove box for calculating production rates (Fig. 4). The films on mica substrates were used for AFM measurements. We also ran the reference experiment in which we flowed the heated gas mixture with $H_2S$ (800 K) through the chamber for 72 hr but did not turn on AC glow discharge or UV lamp. The gas phase composition was monitored by the RGA during the reference experiment. The recorded mass spectra showed no significant intensity changes of the peaks, indicating that heating alone did induce observable chemical modification of the gas composition.

**Atomic force microscopy (AFM) measurements.** We examined the films using a Bruker Dimension 3100 atomic force microscope under ambient conditions (298 K).[8,21] A supersharp AFM silicon probe (SHR150, Budget Sensors; radius: less than 1 nm; cone angle: less than 20°) was used for high-resolution imaging of the surface of the films (Fig. 2). The surfaces were imaged with tapping mode to protect the surface and overcome lateral forces. The haze films on mica discs were used for AFM measurements because the surface of the cleaved mica is molecularly smooth. The smooth surface makes it easier to acquire high resolution AFM images. The AFM images were used to obtain the size distribution of the solid particles (the measured diameter errors are less than 3 nm).[21] To better inspect the size distribution of the haze particles, a larger scanning area (10 μm × 10 μm) of each film was analyzed, and the number of particles used for the size-distribution statistics were 4611 for the UV-no-$H_2S$ experiment, 135610 for the plasma-no-$H_2S$ experiment,17238 for the UV-$H_2S$ case, and 21472 for the plasma-$H_2S$ case. From the AFM images, we counted the number of particles in different sizes and plotted the percentage of particles (N/Ntotal×100%) in 5 nm size bins (Fig. 3).[8,21]

Because the UV experiments did not produce enough solid to collect and weigh, we estimated the production rate based on the size distribution.[8] By assuming that the particle distribution is uniform on the inside wall of the chamber, we calculated the total volume ($V$) of the particles:

$$V = \sum_{i=0}^{i} \frac{\pi}{6} D_i^3 N_i \tag{1}$$

where $D_i$ is the median particle diameter in each bin and $N_i$ is the number of particles in each bin over the total available surface area within the chamber. We assume there is only one layer of the particles on the surface, which means we are obtaining a lower limit: multiple layers of particles are possible. Then, we estimated the mass (mg) and the production rate (mg h$^{-1}$) of the particles if we assume the particle density is the same as a Titan tholin sample ($\rho$ = 1.38 g cm$^{-3}$).[29] Previous studies showed that the particle density varies with the initial gas mixture,[29,51] and the particle density from our experiments could be different between gas mixtures. However, assuming the same density allows us to achieve production rate approximations for the UV experiments and therefore to compare them with those from the plasma experiments (Fig. 4). We calculated the production rates for the plasma experiments in the same way, which are 0.13 mg h$^{-1}$ for the plasma-no-$H_2S$ experiment and 0.25 mg h$^{-1}$ for the plasma-$H_2S$ experiment. These values are lower than those calculated from the weighing method because the plasma experiments produced enough particles to form multiple layers on the surface.

**Gas phase composition measurements.** For each experiment, we measured the composition change of the gas mixture using an RGA (a quadrupole mass spectrometer). An electron ionization (EI) source with standard 70 eV energy was used in the measurements and the scanning mass range was 1-100 amu. The minimum detection limit of the RGA was 0.1 ppm under our operating conditions. We took 50 mass scans of the initial gas mixtures before turning on the energy source (plasma or UV). After turning on the energy source, we waited 30 minutes for the gas mixtures to reach a steady state and then took 1000 mass scans of the gas mixture for the duration (72 h) of each experiment (the gas mixture contains newly-formed gas products and remaining initial gases). We averaged the mass scans to lower the noise level and obtained the average mass spectra of the gas mixtures before and during experiments. The intensity variation for the mass peaks during the 1000 scans at steady state are within 1.0%. The background of the RGA chamber was measured before each experiment and subtracted from the measured spectra of the gas mixtures. We used the total mass intensity to normalize the mass spectra (Fig. 5) for ease of comparison.

By comparing the normalized mass spectra, we identified the mass peaks that decrease or increase significantly (>10%) after turning on the energy source (Table 1). **The relative intensity change of the changed molecular peaks are listed in Table 1.** The peaks that increased are associated with the new gas products. It is difficult to quantify the yield of the gas products because different species have different ionization efficiencies and instrumental responses in the mass spectrometer. If we assume that the molecular peaks in the mass spectra are quantitative and they follow the same linear calibration curves on the RGA mass spectrometer, the increased intensity can represent the relative yield of each species. We grouped the gas products in three categories, including oxygen molecule ($O_2$), organics (organic gas products that contain C, H, O, and N) and sulfur products (both organic and inorganic products that contains sulfur). The relative yield for each category (Fig. 6) was calculated by adding up the increased intensity of the products in the category.

**Additional References**